\documentclass[aps,twocolumn,prl]{revtex4}

\pdfoutput=1

\usepackage[utf8]{inputenc}
\usepackage{lmodern}
\usepackage{hyperref}
\usepackage{amsmath} 
\usepackage{graphicx}
\usepackage{amssymb}
\usepackage{amsthm}
\usepackage{epstopdf} 
\usepackage{natbib}
\usepackage{dcolumn}% Align table columns on decimal point
\usepackage{amsfonts}

\newcommand{\micron}{\,\mathrm{\mu m}}
\newcommand{\cc}{\,\mathrm{cm^{-3}}}
\newcommand{\mJcm}{\,\mathrm{mJ/cm}^2}

\newcommand{\nm}{\,\mathrm{nm}}
\newcommand{\ps}{\,\mathrm{ps}}
\newcommand{\fs}{\,\mathrm{fs}}
\newcommand{\kev}{\,\mathrm{keV}}
\newcommand{\K}{\,\mathrm{K}}

\begin{document}

%\title{Observation of photo-induced dynamical diffraction effects in ultrafast electron diffraction}
%On the role of multiple scattering in ultrafast electron diffraction experiments\\
%Observation of photo-induced multiple scattering in ultrafast electron diffraction
\title{Observation of large multiple scattering effects\\ 
in ultrafast electron diffraction on single crystal silicon}

\author{I. Gonzalez Vallejo,$^{1,2}$ G. Gall\'e,$^1$ B. Arnaud,$^3$ S.~A. Scott,$^4$ M.~G. Lagally,$^4$ D. Boschetto,$^1$ P.E. Coulon,$^5$ G. Rizza,$^5$ F. Houdellier,$^6$ D. Le Bolloc'h,$^2$ and J. Faure$^1$}

\affiliation{$^1$LOA, ENSTA ParisTech, CNRS, Ecole polytechnique, Univ. Paris-Saclay, Palaiseau, France}
\affiliation{$^2$LPS, CNRS, Univ. Paris-Sud, Univ. Paris-Saclay, 91405 Orsay, France}
\affiliation{$^3$Institut des Mol\'ecules et Mat\'eriaux du Mans, UMR CNRS 6283, Le Mans Univ. , 72085 Le Mans, France}
\affiliation{$^4$University of Wisconsin-Madison, Madison, Wisconsin 53706, USA}
\affiliation{$^5$LSI, CNRS, CEA-DRF-IRAMIS, Ecole Polytechnique, Univ. Paris-Saclay, Palaiseau, France}
\affiliation{$^6$CEMES, CNRS, 29 Rue Jeanne Marvig, 31055 Toulouse, France}

\begin{abstract}
We report on ultrafast electron diffraction on high quality single crystal silicon. The ultrafast dynamics of the Bragg peaks exhibits a giant photo-induced response which can only be explained in the framework of dynamical diffraction theory, taking into account multiple scattering of the probing electrons in the sample. In particular, we show that lattice heating following photo-excitation can cause an unexpected increase of the Bragg peak intensities, in contradiction with the well-known Debye-Waller effect. We anticipate that multiple scattering should be systematically considered in ultrafast electron diffraction on high quality crystals as it dominates the Bragg peak dynamics. In addition, taking into account multiple scattering effects opens the way to quantitative studies of non-equilibrium dynamics of defects in quasi-perfect crystals.
\end{abstract}

\maketitle

The field of ultrafast dynamics in condensed matter has been very active in the past decades. Its main motivation is to gain new insight on the complex interplay between the various degrees of freedom in materials (charge, lattice, spins) directly in the time domain. In particular, ultrafast X-ray diffraction  \cite{fritz07,mankowsky2014} and ultrafast electron diffraction (UED) \cite{zewa06,scia11,miller14} are ideal techniques for obtaining valuable information on structural dynamics at the atomic scale. The use of ultrafast electron diffraction in pump-probe experiments has proven to be very efficient for studying the dynamics of photo-induced phase transitions by measuring the relative changes of the diffraction pattern following photo-excitation \cite{baum07,eich10,scia09,gao13,Morrison2014}. Indeed, in the case of a structural phase transition, interpreting the dynamics of the diffraction pattern is relatively straightforward: the change in the crystal symmetry can be monitored through the appearance/disappearance of Bragg peaks \cite{eich10,Morrison2014}. However, a wealth of additional information is contained in the diffraction pattern, \textit{e.g.} lattice heating can be estimated through the change of the Bragg peak intensity due to the Debye-Waller effect. Quantitative analysis mostly relies on the use of kinematical diffraction theory, which assumes that the scattering potential of the crystal lattice is a small perturbation, so that the probing electrons undergo a single elastic scattering event, leading to a weak diffracted intensity compared to the incident electron beam. This theory gives satisfactory results when applied to the case of polycrystalline samples where the grain size is only a few nanometers \cite{harb06,Morrison2014}. It led to quasi-direct measurements of the lattice temperature with sub-picosecond resolution in several materials \cite{harb06,miller14}. However, as high quality single crystal samples adapted to UED experiments are becoming available, kinematical theory does not appear sufficient to explain all experimental results. Several UED studies on high quality crystals, such as silicon \cite{harb09} and graphite \cite{lahme14} have reported large photo-induced changes of the Bragg peak intensity that cannot be explained by kinematical theory. The authors proposed that multiple scattering of the electrons must be at play but no quantitative analysis was performed to fully confirm this hypothesis. 

In electron microscopy, multiple scattering is taken into account in the framework of dynamical diffraction theory \cite{reimer08,fultz13}. In high quality crystals, multiple scattering needs to be considered due to the very high elastic scattering cross section of electrons. Despite this, little attention has been given to these effects in time-resolved electron diffraction experiments. To our knowledge, multiple scattering was considered in detailed only in \cite{scha2011} in a UED experiment in reflexion geometry designed to study surface dynamics. In this letter, we show that multiple scattering completely dominates the dynamics of the diffraction pattern in the commonly used transmission geometry. The experiment is performed on nano-membranes of monocrystalline silicon which is the archetypal example of the perfect single crystal. In addition, the availability of the silicon scattering potential enables a thorough and quantitative comparison between experiment and theory, leading to the unambiguous conclusion that the observed dynamics is dominated by the photo-induced changes of multiple scattering physics.

The electron bunches are first generated by back-illuminating a gold photo-cathode with a $\lambda=266\nm$ ultrashort laser pulse of $<60\fs$ duration. Electrons are then accelerated in a DC gun, delivering accelerating voltages up to $100\kev$, and then focused by a solenoid to a spot size of $150\micron$ Full Width Half Maximum (FWHM) at the sample position. The charge of the electron bunch beam is $<1\;\mathrm{fC}$ resulting in space charge dominated bunches with $<300\fs$ duration. Unless stated otherwise, the electron energy is $45\kev$. The silicon sample is pumped with a $35\fs$ pump laser pulse, with $\lambda=400\nm$ photons. The incident fluence is $12\pm1\mJcm$, over a $~500\micron$ FWHM laser spot. The diffracted peaks are detected with a MCP detector imaged onto a CCD camera. The experiment is performed at 1~kHz repetition rate and each diffraction image is obtained by accumulating over $5000$ pulses. The silicon samples were thinned out from a silicon on insulator wafer \cite{scott07}, resulting in a grid of $350\times350\micron$ free standing nano-membranes with [001] orientation. The membrane thickness was measured using convergent beam electron diffraction \cite{allen81} and estimated to be $70\pm2\nm$. 

We start by reviewing some properties of silicon and its expected dynamical response following photo-excitation. We measured the pump pulse absorption in the sample to be $55\pm5\%$. Thus, starting from an incident fluence of $F_{inc}=12\mJcm$, the absorbed fluence is estimated at $F_{abs}=6.5\mJcm$. The pump laser pulse causes the excitation of electron-hole pairs and the density of excited electrons in the conduction band is given by:  $n_{exc}=F_{abs}/L\hbar\omega$, \textit{i.e.} $n_{exc}=1.8\times10^{21}\cc$ for our experimental parameters.
Excited carriers thermalize via electron-electron scattering on the	100 fs time scale \cite{jeong96} and form two subsystems comprising hot electrons and holes. The electrons (holes) subsequently relax to the bottom of the conduction band 	(top of the valence band) through electron-phonon coupling on a picosecond timescale, causing lattice heating \cite{shank83,harb06}. Using \textit{ab initio} calculations \cite{gonz09} for determining the quasiparticle density of states of the valence and conduction bands \cite{arn05}, as well as the specific heat $C_p(T)$ of silicon, we were able to determine the lattice temperature after electron relaxation assuming that the number of electron-hole pairs stays constant during this part of the dynamics. This gives a lattice temperature increase of $\Delta T=240\K$. Additional delayed heating occurs via electron-hole pair recombination across the gap. At this excitation level, it is well-known that the dominant mechanism is Auger recombination \cite{richter12}. The dynamics of excited carrier is governed by the following equation $dn_{exc}/dt=-(C_e+C_h)n_{exc}^3$, where $C_e$ and $C_h$ are the Auger coefficients for electron and holes respectively. Following Dziewior and Schmid \cite{dzie1977} , we used $C_e+C_h=3.8\times10^{-31}\;\mathrm{cm^6s^{-1}}$, and we find that $90\%$ of the Auger recombination has occurred after $40\ps$ and $94\%$ after $100\ps$. Therefore, after $100\ps$, we estimate a temperature increase of $\Delta T=460\K$. At this point, the system reaches a metastable state as heat diffusion occurs on the microsecond time scale for our sample geometry. In kinematical diffraction, lattice heating manifests itself by the decrease of the Bragg peak intensities according to the Debye-Waller factor: $I_{hkl}(T)=I_{hkl}(0)e^{-2M}$, with $2M=\langle u^2 \rangle\Delta k_{hkl}^2$. Here, $\langle u^2 \rangle$ represents the rms displacement of atoms around their equilibrium position and $\Delta k_{hkl}=4\pi\sin\theta_{hkl}/\lambda$, where $\lambda$ is the electron de Broglie wavelength. Using \textit{ab initio} calculations \cite{arn05,gonz09,lee95} for estimating the values of $\langle u^2 \rangle$, we find that the (220) peaks should all decrease by $10\%$ after lattice heating is completed: $I_{220}(800\K)/I_{220}(300\K)-1=0.9$. This scenario and the use of kinematical theory to interpret the decrease of the Bragg peak intensities was validated in a UED experiment on polycrystalline silicon \cite{harb06}. %$$
%\phi_d(\mathbf{\Delta k},T)\propto \mathcal{S}(\mathbf{\Delta k})\mathcal{F}(\mathbf{\Delta k})e^{-M(T)}
%$$
%hef $\Delta k=$ is the scattering vector, $\mathcal{S}(\mathbf{\Delta k})$ the shape factor (cubic in the case of silicon)

\begin{figure}[t!]
\centerline{\includegraphics[width=8cm]{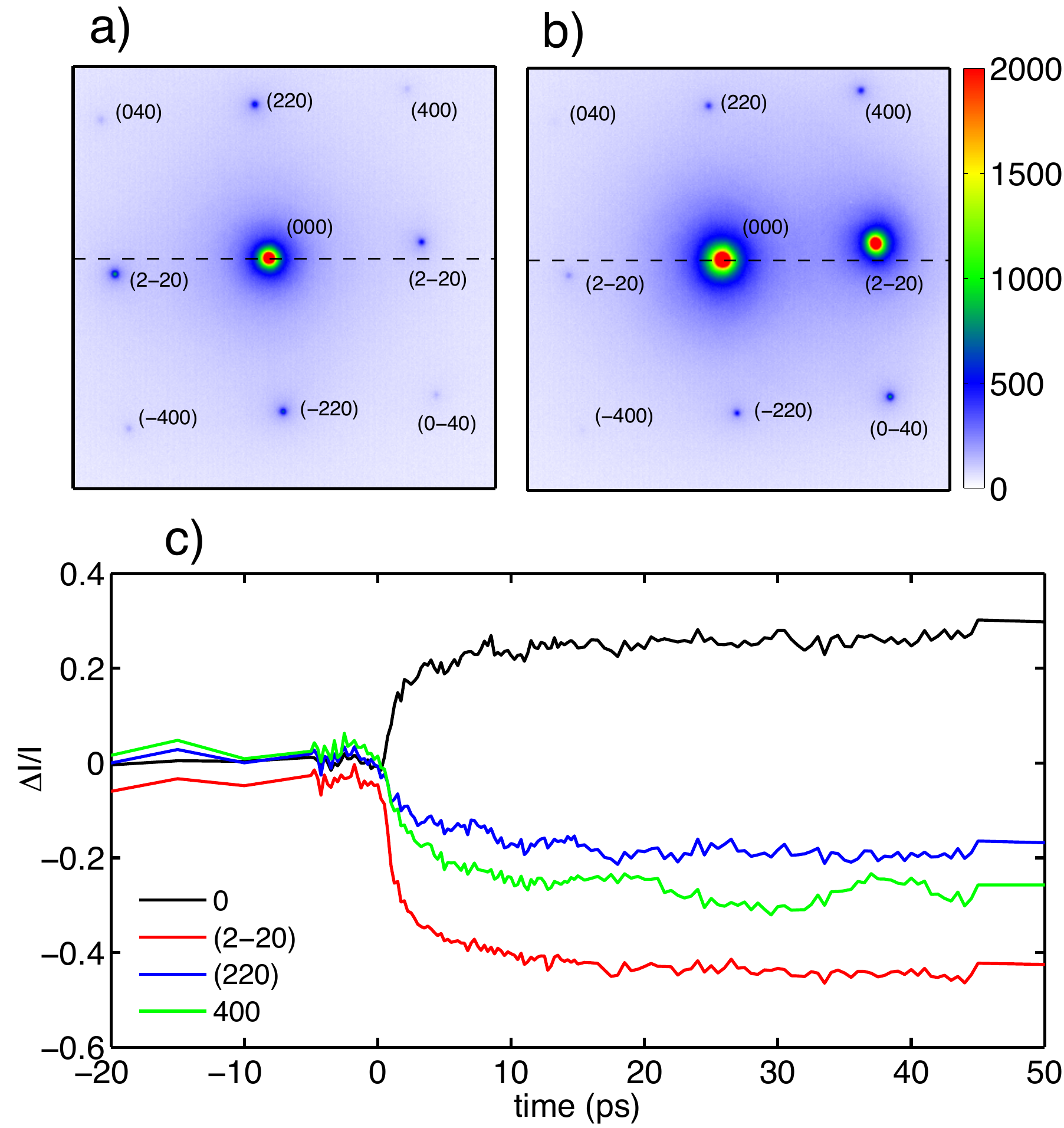}}
\caption{a-b): Diffraction images from a $[001]$ oriented silicon nano-membrane. a) The crystal is oriented such that the electron beam is parallel to the $[001]$ axis. b) The crystal is tilted by the Bragg angle $\theta_{220}=0.84^\circ$ such that the Bragg condition is satisfied for the (2-20) peak. c) Result of a pump-probe scan showing the relative intensity changes of various Bragg peaks $\Delta I/I$. The incident fluence is $12\mJcm$}\label{fig1}
\end{figure}

We now demonstrate that this interpretation does not hold in the case of high quality single crystals. Typical diffraction patterns from the silicon nano-membranes are shown in Fig.~\ref{fig1}. In a), the electron beam is oriented so that it is parallel to the $[001]$ direction: the diffraction pattern is symmetric and the various (220) peaks have similar intensity. The diffracted beam intensities is about one order of magnitude lower compared to the intensity of the transmitted electron beam (referred to as the 0-order beam in the following). In contrast, in b) the sample was tilted along the horizontal axis (represented by the dashed black line) so that the (2-20) peak satisfies the Bragg condition. The diffraction pattern is quite asymmetric and remarkably, the 0-order and the (2-20) peak have similar intensities. This fact clearly contradicts the basic hypothesis of kinematical diffraction theory which states that the diffracted intensity is much lower than the transmitted beam intensity. Fig.~\ref{fig1}c shows the dynamics of various Bragg peaks following photo-excitation at incident fluence of $12\mJcm$. In this case, the sample was oriented so that the (2-20) peak is slightly off Bragg. All Bragg peaks exhibit similar dynamics: the relative intensity $\Delta I/I$ starts with a sharp decrease on the picosecond time scale. This is followed by a slower roll-off and further decrease on the 10 ps time scale. According to the above-mentioned scenario, the fast picosecond time scale can be attributed to electron relaxation and lattice heating via electron phonon coupling while the slower time scale can be attributed to delayed heating due to Auger recombination. After tens of picoseconds, the Bragg peak intensity is relatively flat and a quasi-steady state is established that lasts hundreds of picoseconds.

\begin{figure}[t!]
\centerline{\includegraphics[width=8cm]{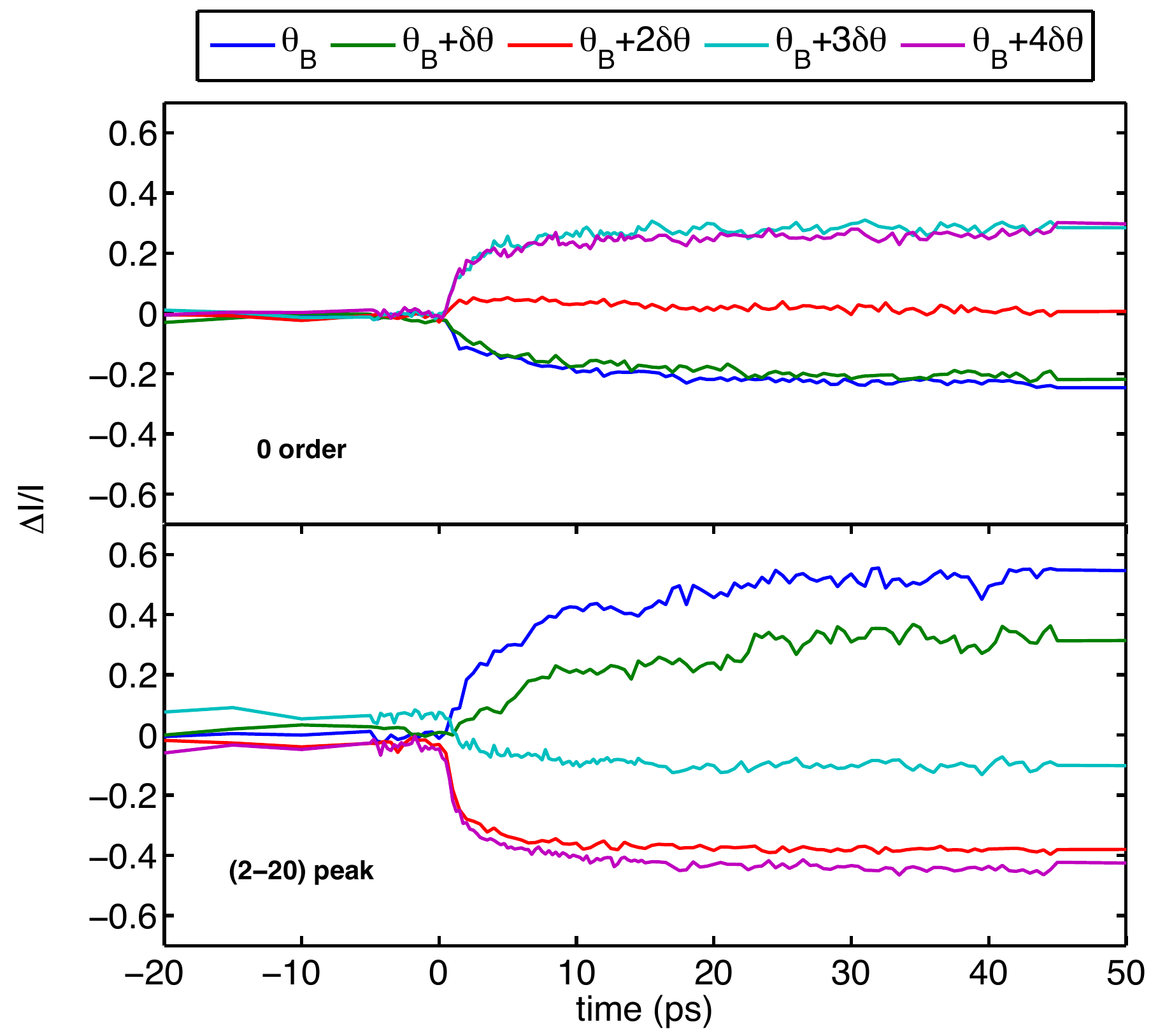}}
\caption{Photo-induced dynamics for various sample orientations. The blue curve is obtained when the sample is exactly at the Bragg angle; the other curves are obtained by tilting the sample by $\delta \theta=0.26^\circ$. The incident fluence is $12\mJcm$. Top: dynamics of the transmitted beam. Bottom: dynamics of the (2-20) peak.}\label{fig2}
\end{figure}

These different time scales are consistent with previous results \cite{harb09}, but a truly intriguing feature is the magnitude of the measured signal: the (2-20) peak decreases by $40\%$ while the 0-order peak increases by nearly $30\%$. Even more surprisingly, we observed that the dynamics of the Bragg peak is extremely sensitive to sample orientation. In Fig.~\ref{fig2}, we show the dynamics $\Delta I/I (t)$ for the transmitted beam (a) and for the (2-20) peak (b) for five different sample orientations.  The results are striking as a $~1^\circ$ tilt can turn the intensity change of the (2-20) peak from $-40\%$ to almost $60\%$. Therefore, we not only observe a giant photo-induced response in the Bragg peak intensity but the sign of the response $\Delta I/I$ is determined by sample orientation. It is also interesting to note that the 0-order and the (2-20) peak have a complementary behavior, indicating a possible coupling.

These observations are in complete contradiction with the predictions of kinematical theory. In kinematical theory, the 0-order should remain unchanged while the all (220) peaks should decrease by less than $10\%$. Finally, the magnitude of the intensity changes $\Delta I/I$ should be independent on sample orientation.

\begin{figure}[t!]
\centerline{\includegraphics[width=7cm]{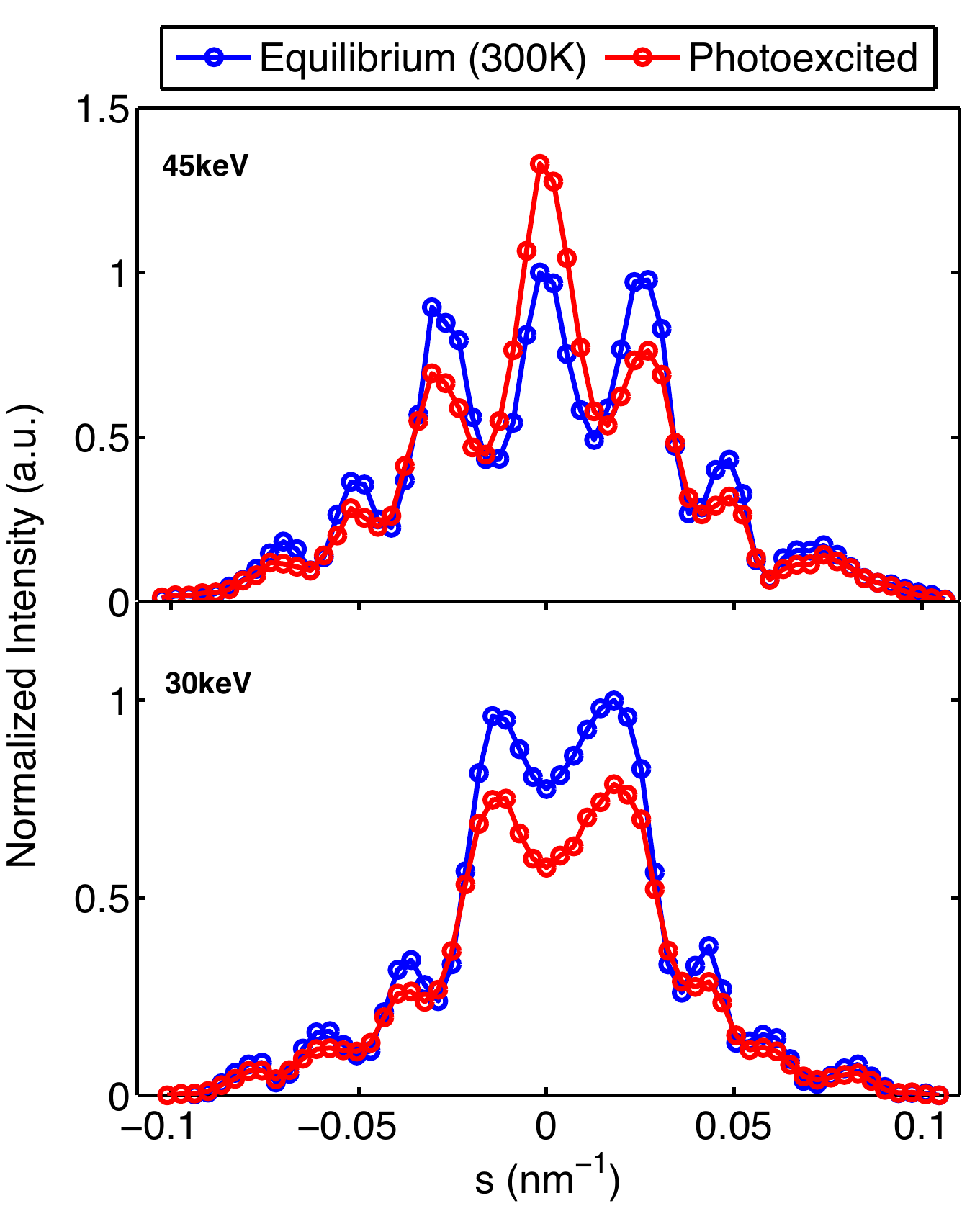}}
\caption{Top: experimental rocking curves for the (2-20) peak taken with $45\kev$ electrons, at equilibrium $T=300\K$ (blue curve) and in the photo-excited state (red curve), taken at $t=150\ps$ delay. Bottom: same but using $30\kev$ electrons as a probe. The rocking curves were normalized relative to the equilibrium case.}\label{fig3}
\end{figure}

In order to gain further insight on these large changes of intensity, we measured the rocking curves of several diffraction peaks. Figure \ref{fig3} shows the rocking curve of the (2-20) peak at equilibrium (i.e. at $300\K$, blue curves) and in the photo-excited state (red curves) taken 150~ps after the arrival of the pump pulse, \textit{i.e.} after thermalization of the sample has occurred. Rocking curves are shown at two different electron energies. We plot the Bragg peak intensity $I(s)$, where $s$ is the amplitude of the deviation vector $\mathbf{s}=\mathbf{\Delta k}-\mathbf{g}$, and $\mathbf{g}$ is the lattice reciprocal vector corresponding to the (2-20) peak. Figure \ref{fig3} clearly shows that the shape and magnitude of the rocking curve changes upon photo-excitation. However, there is no angular shift of the rocking curve upon photo excitation, invalidating previous interpretations based on lattice expansion \cite{lahme14} or sample distortion \cite{harb09}. In addition, the results of Fig.~\ref{fig3}  summarize and clarify the surprising features of Fig.~\ref{fig2}: for $45\kev$ electrons, the intensity change is positive at the Bragg angle, whereas it is negative for most off-Bragg cases. For $30\kev$ electrons, the behavior is quite different: here, the intensity change is always negative after photo-excitation. The shape of these rocking curves, by departing from the usual $\sin^2x/x^2$ line shape of kinematical theory, indicates that dynamical effects are dominating the physics of electron diffraction, even at equilibrium.

\begin{figure}[t!]
\centerline{\includegraphics[width=7cm]{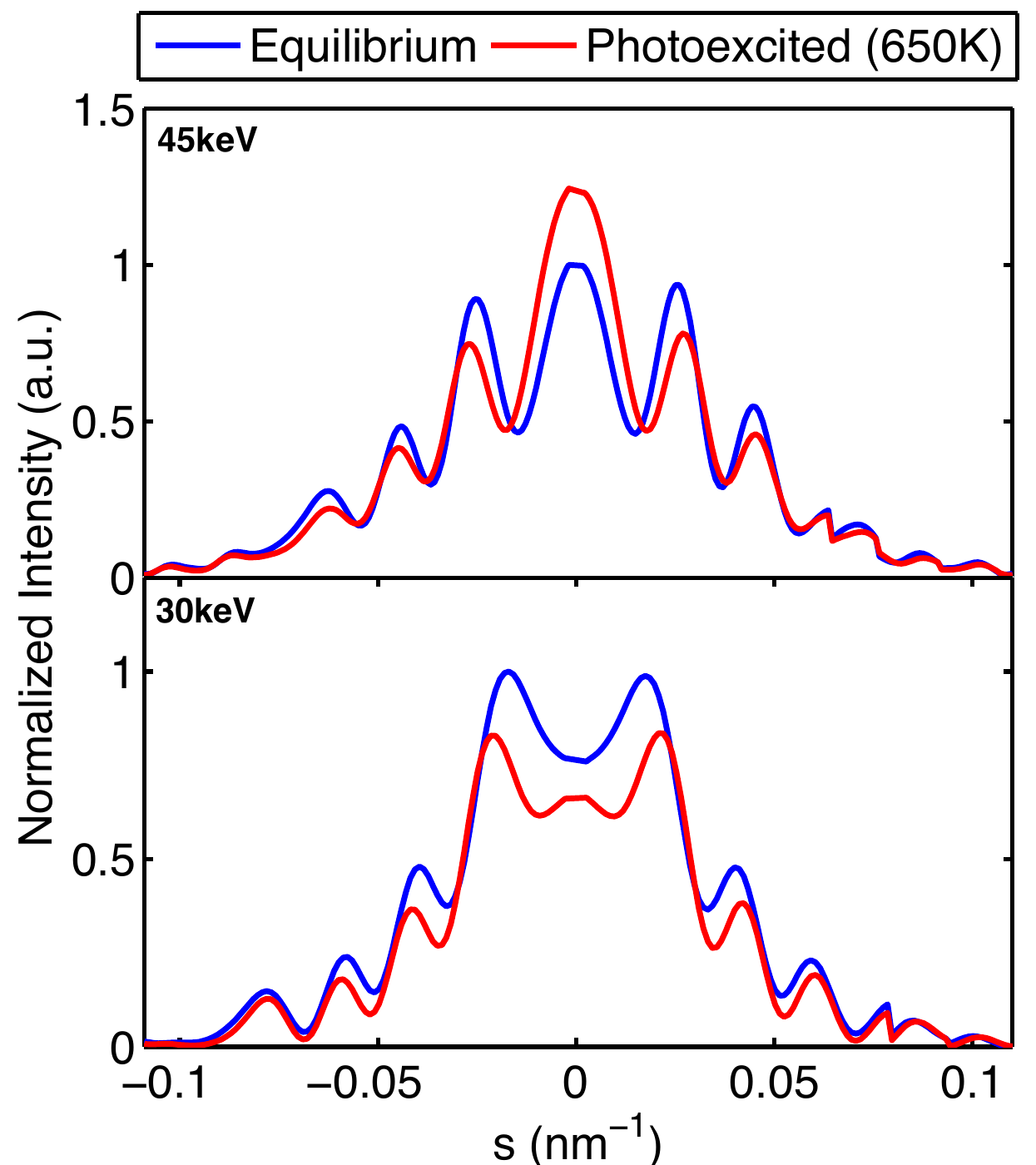}}
\caption{Results of N-beam dynamical diffraction theory with $N=26$ beams. Top: calculated rocking curves for the (2-20) peak  in the case of $45\kev$ electrons. The blue curve shows the result at $T=300\K$ and at $T=650\K$ (red curve). Bottom: same calculations but with $30\kev$ electrons. A gaussian background was added to the N-beam calculations in order to better fit the data. The rocking curves are normalized relative to the equilibrium case.}\label{fig4}
\end{figure}

The fact that the rocking curve changes with temperature and electron energy can be understood quantitatively using a simplified version of dynamical diffraction theory: the 2-beam theory where one considers only the transmitted beam and one diffracted beam with intensity $I_{\mathbf{g}}$. In 2-beam theory, the diffracted intensity depends on the thickness of the sample $L$ and reads
\begin{equation}
I_\mathbf{g}(s,L)=\frac{1}{V}\frac{\sin^2(s_{e} L/2)}{(s_{e}\xi_\mathbf{g})^2}  
\end{equation}
where $s_e=\sqrt{s^2+1/\xi_\mathbf{g}^2}$ is the amplitude of the effective deviation vector and $\xi_\mathbf{g}$ is the extinction distance. 
%At the Bragg angle, $s=0$, the diffracted intensity evolves during propagation in the sample as $I_\mathbf{g}(t)\propto \sin^2(t/2\xi_\mathbf{g})$. Thus, the Bragg peak reaches a maximum at $t=\pi\xi_\mathbf{g}$ and cancels out at $t=2\pi\xi_\mathbf{g}$ ($=56\nm$ is silicon at $45\kev$). Therefore, the extinction distance is responsible the intensity of the diffracted peak along propagation. 
The extinction distance defines the shape of the rocking curve and changes of $\xi_\mathbf{g}$ will modify the rocking curve. 

At $T=0\K$, the extinction distance reads $\xi_\mathbf{g}=\frac{1}{\gamma\lambda}\frac{\pi\hbar^2}{m_eU_g}$
where $m_e$ is the electron mass, $\gamma=1+E/m_ec^2$ is the Lorentz factor of an electron with kinetic energy $E$. The two beams are coupled through $U_\mathbf{g}$, the Fourier component of the crystal potential $V(\mathbf{r})$ corresponding to reciprocal lattice vector $\mathbf{g}$: $V(\mathbf{r})=\sum_\mathbf{g}U_\mathbf{g}e^{i\mathbf{g}\cdot\mathbf{r}}$. Clearly, the extinction distance depends on electron energy via $\gamma\lambda$, explaining why the rocking curve changes with electron energy. The temperature dependence can be accounted for by formally replacing $U_\mathbf{g}$ by $U_\mathbf{g}e^{-M}$ \cite{takagi58}. Consequently, the extinction distance increases with temperature  \cite{thomas1965} like $\xi_\mathbf{g}(T)=\xi_\mathbf{g}(0)e^{M}$. Evidently, a rise in temperature causes an increase of $\xi_\mathbf{g}$, implying changes of the shape of the rocking curve.

We found that 2-beam theory does not allow us to fit our experimental rocking curves and that additional Bragg peaks need to be taken into account. This is also apparent in the experimental data of Fig.~\ref{fig2}: the diffracted intensity is not conserved if one considers only the 0-order and the (2-20) peak, indicating that more diffracted beams need to be considered. Therefore, we turned to a N-beam theory and solved the Howie-Whelan equations \cite{howie61}
\begin{equation}
\frac{\partial\phi_{\mathbf{g}}}{\partial z}=is_{\mathbf{g}}\phi_{\mathbf{g}}+\sum_{\mathbf{g'}\neq\mathbf{g}}\frac{i}{2\xi_{\mathbf{g-g'}}}\phi_{\mathbf{g'}}
\end{equation}
Here, $\phi_{\mathbf{g}}$ is the amplitude of the diffraction peak $\mathbf{g}$ and two peaks $\phi_{\mathbf{g}}$ and $\phi_{\mathbf{g'}}$ are coupled through the extinction distance $\xi_{\mathbf{g-g'}}\propto 1/U_\mathbf{g-g'}$. Implementing this method requires the detailed knowledge of the scattering potential. Silicon data on the various $U_\mathbf{g-g'}$ was taken from the code JEMS \cite{JEMS}. In the experiment, we detect 12 diffraction peaks during a rocking curve scan but we found that the N-beam theory converges for $N>24$ and we present results with $N=26$ (more details can be found in the suppl. info). We found that N-beam theory reproduces all the trends of the experiment but the data could be even better fitted by adding a gaussian background to the results of the N-beam calculations. This background is a phenomenological modeling of absorption and inelastic scattering effect. The blue curves in Fig.~\ref{fig4} show the rocking curves at $300\K$, both for $45\kev$ and $30\kev$ electrons. The photo-excited state was best fitted considering a $T=650\K$ temperature. Results are represented by the red curves in Fig.~\ref{fig4}, showing excellent agreement with the measurements. We conclude that the observed dynamics of the Bragg peaks and in particular the behavior of $\Delta I/I$ can be fully explained by lattice heating and dynamical diffraction effects. In particular, we obtained the non-intuitive result that depending on the electron energy and the sample orientation, lattice heating can cause an increase of the Bragg peak intensity, contrary to the well-known Debye-Waller effect.

While this study was performed on silicon, we anticipate that such effects should be present in all materials provided that the crystal quality is high and the thickness comparable with the extinction distance. Indeed, when $L\ll 2\pi\xi_{\mathbf{g}}$, multiple scattering can be neglected and kinematic theory appears to be a valid approximation. Typical extinction distances are tens of nanometers ($2\pi\xi_{\mathbf{220}}=56\nm$ for silicon at $45\kev$), so that multiple scattering and dynamical effects have to be considered as soon as the sample thickness is larger that a 1-10~nm, depending on the material. Consequently, when high quality single crystals are used, the quantitative interpretation of UED experiments might become quite complex as modeling multiple scattering requires prior knowledge of the crystal scattering potential. Dynamical effects, in turn, could potentially be used to obtain new information on the dynamics of the crystal potential. Finally, dynamical effects are also useful to visualize crystal defects, such as dislocation or stacking faults \cite{fultz13}. Therefore, they should enable a new type of experiments in which the dynamics of defects following laser irradiation can be studied using ultrafast electron imaging.

\begin{acknowledgments} 
Acknowledgments: This work was funded by the European Research Council under Contract No. 306708, ERC Starting Grant FEMTOELEC and also by a public grant from the “Laboratoire d’Excellence Physics Atoms Light Mater”,  LabEx PALM, Contract No. ANR-11-IDEX-0003-02. Fabrication and characterization of Si nano-membrane samples (S.A.S. and M.G.L.) was supported by the US Department of Energy Grant No. DE-FG0203ER46028.
\end{acknowledgments}

%\bibliographystyle{apsrev}
%\bibliography{jerome_SS_1017}

\begin{thebibliography}{29}
\expandafter\ifx\csname natexlab\endcsname\relax\def\natexlab#1{#1}\fi
\expandafter\ifx\csname bibnamefont\endcsname\relax
  \def\bibnamefont#1{#1}\fi
\expandafter\ifx\csname bibfnamefont\endcsname\relax
  \def\bibfnamefont#1{#1}\fi
\expandafter\ifx\csname citenamefont\endcsname\relax
  \def\citenamefont#1{#1}\fi
\expandafter\ifx\csname url\endcsname\relax
  \def\url#1{\texttt{#1}}\fi
\expandafter\ifx\csname urlprefix\endcsname\relax\def\urlprefix{URL }\fi
\providecommand{\bibinfo}[2]{#2}
\providecommand{\eprint}[2][]{\url{#2}}

\bibitem[{\citenamefont{Fritz et~al.}(2007)\citenamefont{Fritz, Reis, Adams,
  Akre, Arthur, Blome, Bucksbaum, Cavalieri, Engemann, Fahy et~al.}}]{fritz07}
\bibinfo{author}{\bibfnamefont{D.~M.} \bibnamefont{Fritz}},
  \bibinfo{author}{\bibfnamefont{D.~A.} \bibnamefont{Reis}},
  \bibinfo{author}{\bibfnamefont{B.}~\bibnamefont{Adams}},
  \bibinfo{author}{\bibfnamefont{R.~A.} \bibnamefont{Akre}},
  \bibinfo{author}{\bibfnamefont{J.}~\bibnamefont{Arthur}},
  \bibinfo{author}{\bibfnamefont{C.}~\bibnamefont{Blome}},
  \bibinfo{author}{\bibfnamefont{P.~H.} \bibnamefont{Bucksbaum}},
  \bibinfo{author}{\bibfnamefont{A.~L.} \bibnamefont{Cavalieri}},
  \bibinfo{author}{\bibfnamefont{S.}~\bibnamefont{Engemann}},
  \bibinfo{author}{\bibfnamefont{S.}~\bibnamefont{Fahy}}, \bibnamefont{et~al.},
  \bibinfo{journal}{Science} \textbf{\bibinfo{volume}{315}},
  \bibinfo{pages}{633} (\bibinfo{year}{2007}).

\bibitem[{\citenamefont{Mankowsky et~al.}(2014)\citenamefont{Mankowsky, Subedi,
  F{\"o}rst, Mariager, Chollet, Lemke, Robinson, Glownia, Minitti, Frano
  et~al.}}]{mankowsky2014}
\bibinfo{author}{\bibfnamefont{R.}~\bibnamefont{Mankowsky}},
  \bibinfo{author}{\bibfnamefont{A.}~\bibnamefont{Subedi}},
  \bibinfo{author}{\bibfnamefont{M.}~\bibnamefont{F{\"o}rst}},
  \bibinfo{author}{\bibfnamefont{S.~O.} \bibnamefont{Mariager}},
  \bibinfo{author}{\bibfnamefont{M.}~\bibnamefont{Chollet}},
  \bibinfo{author}{\bibfnamefont{H.~T.} \bibnamefont{Lemke}},
  \bibinfo{author}{\bibfnamefont{J.~S.} \bibnamefont{Robinson}},
  \bibinfo{author}{\bibfnamefont{J.~M.} \bibnamefont{Glownia}},
  \bibinfo{author}{\bibfnamefont{M.~P.} \bibnamefont{Minitti}},
  \bibinfo{author}{\bibfnamefont{A.}~\bibnamefont{Frano}},
  \bibnamefont{et~al.}, \bibinfo{journal}{Nature}
  \textbf{\bibinfo{volume}{516}}, \bibinfo{pages}{71} (\bibinfo{year}{2014}).

\bibitem[{\citenamefont{Zewail}(2006)}]{zewa06}
\bibinfo{author}{\bibfnamefont{A.~H.} \bibnamefont{Zewail}},
  \bibinfo{journal}{Annu. Rev. Phys. Chem.} \textbf{\bibinfo{volume}{57}},
  \bibinfo{pages}{65} (\bibinfo{year}{2006}).

\bibitem[{\citenamefont{Sciaini and Miller}(2011)}]{scia11}
\bibinfo{author}{\bibfnamefont{G.}~\bibnamefont{Sciaini}} \bibnamefont{and}
  \bibinfo{author}{\bibfnamefont{R.~J.~D.} \bibnamefont{Miller}},
  \bibinfo{journal}{Rep. Prog. Phys.} \textbf{\bibinfo{volume}{74}},
  \bibinfo{pages}{096101} (\bibinfo{year}{2011}).

\bibitem[{\citenamefont{Miller}(2014)}]{miller14}
\bibinfo{author}{\bibfnamefont{R.~J.~D.} \bibnamefont{Miller}},
  \bibinfo{journal}{Science} \textbf{\bibinfo{volume}{343}},
  \bibinfo{pages}{1108} (\bibinfo{year}{2014}).

\bibitem[{\citenamefont{Baum et~al.}(2007)\citenamefont{Baum, Yang, and
  Zewail}}]{baum07}
\bibinfo{author}{\bibfnamefont{P.}~\bibnamefont{Baum}},
  \bibinfo{author}{\bibfnamefont{D.~S.} \bibnamefont{Yang}}, \bibnamefont{and}
  \bibinfo{author}{\bibfnamefont{A.~H.} \bibnamefont{Zewail}},
  \bibinfo{journal}{Science} \textbf{\bibinfo{volume}{318}},
  \bibinfo{pages}{788} (\bibinfo{year}{2007}).

\bibitem[{\citenamefont{Eichberger et~al.}(2010)\citenamefont{Eichberger,
  Sch\"afer, Krumova, Beyer, Demsar, Berger, Moriena, Sciani, and
  Miller}}]{eich10}
\bibinfo{author}{\bibfnamefont{M.}~\bibnamefont{Eichberger}},
  \bibinfo{author}{\bibfnamefont{H.}~\bibnamefont{Sch\"afer}},
  \bibinfo{author}{\bibfnamefont{M.}~\bibnamefont{Krumova}},
  \bibinfo{author}{\bibfnamefont{M.}~\bibnamefont{Beyer}},
  \bibinfo{author}{\bibfnamefont{J.}~\bibnamefont{Demsar}},
  \bibinfo{author}{\bibfnamefont{H.}~\bibnamefont{Berger}},
  \bibinfo{author}{\bibfnamefont{G.}~\bibnamefont{Moriena}},
  \bibinfo{author}{\bibfnamefont{G.}~\bibnamefont{Sciani}}, \bibnamefont{and}
  \bibinfo{author}{\bibfnamefont{R.~J.~D.} \bibnamefont{Miller}},
  \bibinfo{journal}{Nature} \textbf{\bibinfo{volume}{468}},
  \bibinfo{pages}{799} (\bibinfo{year}{2010}).

\bibitem[{\citenamefont{Sciaini et~al.}(2009)\citenamefont{Sciaini, Harb,
  Kruglik, Payer, Hebeisen, Heringdorf, Yamagushi, Hoegen, Ernstorfer, and
  Miller}}]{scia09}
\bibinfo{author}{\bibfnamefont{G.}~\bibnamefont{Sciaini}},
  \bibinfo{author}{\bibfnamefont{M.}~\bibnamefont{Harb}},
  \bibinfo{author}{\bibfnamefont{S.~G.} \bibnamefont{Kruglik}},
  \bibinfo{author}{\bibfnamefont{T.}~\bibnamefont{Payer}},
  \bibinfo{author}{\bibfnamefont{C.~T.} \bibnamefont{Hebeisen}},
  \bibinfo{author}{\bibfnamefont{F.~M.} \bibnamefont{Heringdorf}},
  \bibinfo{author}{\bibfnamefont{M.}~\bibnamefont{Yamagushi}},
  \bibinfo{author}{\bibfnamefont{M.~H.} \bibnamefont{Hoegen}},
  \bibinfo{author}{\bibfnamefont{R.}~\bibnamefont{Ernstorfer}},
  \bibnamefont{and} \bibinfo{author}{\bibfnamefont{R.~J.~D.}
  \bibnamefont{Miller}}, \bibinfo{journal}{Nature}
  \textbf{\bibinfo{volume}{458}}, \bibinfo{pages}{458} (\bibinfo{year}{2009}).

\bibitem[{\citenamefont{Gao et~al.}(2013)\citenamefont{Gao, Lu, Jean-Ruel, Liu,
  Marx, Onda, Koshihara, Nakano, Shao, Hiramatsu et~al.}}]{gao13}
\bibinfo{author}{\bibfnamefont{M.}~\bibnamefont{Gao}},
  \bibinfo{author}{\bibfnamefont{C.}~\bibnamefont{Lu}},
  \bibinfo{author}{\bibfnamefont{H.}~\bibnamefont{Jean-Ruel}},
  \bibinfo{author}{\bibfnamefont{L.~C.} \bibnamefont{Liu}},
  \bibinfo{author}{\bibfnamefont{A.}~\bibnamefont{Marx}},
  \bibinfo{author}{\bibfnamefont{K.}~\bibnamefont{Onda}},
  \bibinfo{author}{\bibfnamefont{S.}~\bibnamefont{Koshihara}},
  \bibinfo{author}{\bibfnamefont{Y.}~\bibnamefont{Nakano}},
  \bibinfo{author}{\bibfnamefont{X.}~\bibnamefont{Shao}},
  \bibinfo{author}{\bibfnamefont{T.}~\bibnamefont{Hiramatsu}},
  \bibnamefont{et~al.}, \bibinfo{journal}{Nature}
  \textbf{\bibinfo{volume}{496}}, \bibinfo{pages}{343} (\bibinfo{year}{2013}).

\bibitem[{\citenamefont{Morrison et~al.}(2014)\citenamefont{Morrison,
  Chatelain, Tiwari, Hendaoui, Bruh{\'a}cs, Chaker, and Siwick}}]{Morrison2014}
\bibinfo{author}{\bibfnamefont{V.~R.} \bibnamefont{Morrison}},
  \bibinfo{author}{\bibfnamefont{R.~P.} \bibnamefont{Chatelain}},
  \bibinfo{author}{\bibfnamefont{K.~L.} \bibnamefont{Tiwari}},
  \bibinfo{author}{\bibfnamefont{A.}~\bibnamefont{Hendaoui}},
  \bibinfo{author}{\bibfnamefont{A.}~\bibnamefont{Bruh{\'a}cs}},
  \bibinfo{author}{\bibfnamefont{M.}~\bibnamefont{Chaker}}, \bibnamefont{and}
  \bibinfo{author}{\bibfnamefont{B.~J.} \bibnamefont{Siwick}},
  \bibinfo{journal}{Science} \textbf{\bibinfo{volume}{346}},
  \bibinfo{pages}{445} (\bibinfo{year}{2014}).

\bibitem[{\citenamefont{Harb et~al.}(2006)\citenamefont{Harb, Ernstorfer,
  Dartigalongue, Hebeisen, Jordan, and Miller}}]{harb06}
\bibinfo{author}{\bibfnamefont{M.}~\bibnamefont{Harb}},
  \bibinfo{author}{\bibfnamefont{R.}~\bibnamefont{Ernstorfer}},
  \bibinfo{author}{\bibfnamefont{T.}~\bibnamefont{Dartigalongue}},
  \bibinfo{author}{\bibfnamefont{C.~T.} \bibnamefont{Hebeisen}},
  \bibinfo{author}{\bibfnamefont{R.~E.} \bibnamefont{Jordan}},
  \bibnamefont{and} \bibinfo{author}{\bibfnamefont{R.~J.~D.}
  \bibnamefont{Miller}}, \bibinfo{journal}{J. Phys. Chem. B}
  \textbf{\bibinfo{volume}{110}}, \bibinfo{pages}{25308}
  (\bibinfo{year}{2006}).

\bibitem[{\citenamefont{Harb et~al.}(2009)\citenamefont{Harb, Peng, Sciaini,
  Hebeisen, Ernstorfer, Eriksson, Lagally, Kruglik, and Miller}}]{harb09}
\bibinfo{author}{\bibfnamefont{M.}~\bibnamefont{Harb}},
  \bibinfo{author}{\bibfnamefont{W.}~\bibnamefont{Peng}},
  \bibinfo{author}{\bibfnamefont{G.}~\bibnamefont{Sciaini}},
  \bibinfo{author}{\bibfnamefont{C.~T.} \bibnamefont{Hebeisen}},
  \bibinfo{author}{\bibfnamefont{R.}~\bibnamefont{Ernstorfer}},
  \bibinfo{author}{\bibfnamefont{M.~A.} \bibnamefont{Eriksson}},
  \bibinfo{author}{\bibfnamefont{M.~G.} \bibnamefont{Lagally}},
  \bibinfo{author}{\bibfnamefont{S.~G.} \bibnamefont{Kruglik}},
  \bibnamefont{and} \bibinfo{author}{\bibfnamefont{R.~J.~D.}
  \bibnamefont{Miller}}, \bibinfo{journal}{Phys. Rev. B}
  \textbf{\bibinfo{volume}{79}}, \bibinfo{pages}{094301}
  (\bibinfo{year}{2009}).

\bibitem[{\citenamefont{Lahme et~al.}(2014)\citenamefont{Lahme, Kealhofer,
  Krausz, and Baum}}]{lahme14}
\bibinfo{author}{\bibfnamefont{S.}~\bibnamefont{Lahme}},
  \bibinfo{author}{\bibfnamefont{C.}~\bibnamefont{Kealhofer}},
  \bibinfo{author}{\bibfnamefont{F.}~\bibnamefont{Krausz}}, \bibnamefont{and}
  \bibinfo{author}{\bibfnamefont{P.}~\bibnamefont{Baum}},
  \bibinfo{journal}{Struc. Dynam.} \textbf{\bibinfo{volume}{1}},
  \bibinfo{pages}{034303} (\bibinfo{year}{2014}).

\bibitem[{\citenamefont{Reimer and Kohl}(2008)}]{reimer08}
\bibinfo{author}{\bibfnamefont{L.}~\bibnamefont{Reimer}} \bibnamefont{and}
  \bibinfo{author}{\bibfnamefont{H.}~\bibnamefont{Kohl}},
  \emph{\bibinfo{title}{Transmission electron microscopy}}
  (\bibinfo{publisher}{Springer Series in Optical Sciences},
  \bibinfo{address}{New York}, \bibinfo{year}{2008}).

\bibitem[{\citenamefont{Fultz and Howe}(2013)}]{fultz13}
\bibinfo{author}{\bibfnamefont{B.}~\bibnamefont{Fultz}} \bibnamefont{and}
  \bibinfo{author}{\bibfnamefont{J.}~\bibnamefont{Howe}},
  \emph{\bibinfo{title}{Transmission electron microscopy and diffractometry of
  materials}} (\bibinfo{publisher}{Springer Series in Graduate Texts in
  Physics}, \bibinfo{address}{Berlin, Heidelberg}, \bibinfo{year}{2013}).

\bibitem[{\citenamefont{Sch{\"a}fer et~al.}(2011)\citenamefont{Sch{\"a}fer,
  Liang, and Zewail}}]{scha2011}
\bibinfo{author}{\bibfnamefont{S.}~\bibnamefont{Sch{\"a}fer}},
  \bibinfo{author}{\bibfnamefont{W.}~\bibnamefont{Liang}}, \bibnamefont{and}
  \bibinfo{author}{\bibfnamefont{A.~H.} \bibnamefont{Zewail}},
  \bibinfo{journal}{J. Chem. Phys.} \textbf{\bibinfo{volume}{135}},
  \bibinfo{pages}{214201} (\bibinfo{year}{2011}).

\bibitem[{\citenamefont{Scott and Lagally}(2007)}]{scott07}
\bibinfo{author}{\bibfnamefont{S.~A.} \bibnamefont{Scott}} \bibnamefont{and}
  \bibinfo{author}{\bibfnamefont{M.~G.} \bibnamefont{Lagally}},
  \bibinfo{journal}{J. of Phys. D: App. Phys.} \textbf{\bibinfo{volume}{40}},
  \bibinfo{pages}{R75} (\bibinfo{year}{2007}).

\bibitem[{\citenamefont{Allen}(1981)}]{allen81}
\bibinfo{author}{\bibfnamefont{F.~S.} \bibnamefont{Allen}},
  \bibinfo{journal}{Phil. Mag. A} \textbf{\bibinfo{volume}{43}},
  \bibinfo{pages}{325} (\bibinfo{year}{1981}).

\bibitem[{\citenamefont{Jeong et~al.}(1996)\citenamefont{Jeong, Zacharias, and
  Bokor}}]{jeong96}
\bibinfo{author}{\bibfnamefont{S.}~\bibnamefont{Jeong}},
  \bibinfo{author}{\bibfnamefont{H.}~\bibnamefont{Zacharias}},
  \bibnamefont{and} \bibinfo{author}{\bibfnamefont{J.}~\bibnamefont{Bokor}},
  \bibinfo{journal}{Phys. Rev. B} \textbf{\bibinfo{volume}{54}},
  \bibinfo{pages}{R17300} (\bibinfo{year}{1996}).

\bibitem[{\citenamefont{Shank et~al.}(1983)\citenamefont{Shank, R.Yen, and
  C.Hirlimann}}]{shank83}
\bibinfo{author}{\bibfnamefont{C.V.}~\bibnamefont{Shank}},
  \bibinfo{author}{\bibnamefont{R.Yen}}, \bibnamefont{and}
  \bibinfo{author}{\bibnamefont{C.Hirlimann}}, \bibinfo{journal}{Phys. Rev.
  Lett.} \textbf{\bibinfo{volume}{50}}, \bibinfo{pages}{454}
  (\bibinfo{year}{1983}).

\bibitem[{\citenamefont{Gonze et~al.}(2009)\citenamefont{Gonze, Amadon,
  Anglade, Beuken, Bottin, Boulanger, Bruneval, Caliste, Caracas, C\^ot\'e
  et~al.}}]{gonz09}
\bibinfo{author}{\bibfnamefont{X.}~\bibnamefont{Gonze}},
  \bibinfo{author}{\bibfnamefont{B.}~\bibnamefont{Amadon}},
  \bibinfo{author}{\bibfnamefont{P.}~\bibnamefont{Anglade}},
  \bibinfo{author}{\bibfnamefont{J.}~\bibnamefont{Beuken}},
  \bibinfo{author}{\bibfnamefont{F.}~\bibnamefont{Bottin}},
  \bibinfo{author}{\bibfnamefont{P.}~\bibnamefont{Boulanger}},
  \bibinfo{author}{\bibfnamefont{F.}~\bibnamefont{Bruneval}},
  \bibinfo{author}{\bibfnamefont{D.}~\bibnamefont{Caliste}},
  \bibinfo{author}{\bibfnamefont{R.}~\bibnamefont{Caracas}},
  \bibinfo{author}{\bibfnamefont{M.}~\bibnamefont{C\^ot\'e}},
  \bibnamefont{et~al.}, \bibinfo{journal}{Computer Physics Communications}
  \textbf{\bibinfo{volume}{180}}, \bibinfo{pages}{2582 }
  (\bibinfo{year}{2009}).

\bibitem[{\citenamefont{Arnaud et~al.}(2005)\citenamefont{Arnaud, Leb\`egue,
  and Alouani}}]{arn05}
\bibinfo{author}{\bibfnamefont{B.}~\bibnamefont{Arnaud}},
  \bibinfo{author}{\bibfnamefont{S.}~\bibnamefont{Leb\`egue}},
  \bibnamefont{and} \bibinfo{author}{\bibfnamefont{M.}~\bibnamefont{Alouani}},
  \bibinfo{journal}{Phys. Rev. B} \textbf{\bibinfo{volume}{71}},
  \bibinfo{pages}{035308} (\bibinfo{year}{2005}).

\bibitem[{\citenamefont{Richter et~al.}(2012)\citenamefont{Richter, Glunz,
  Werner, and J.~Schmidt}}]{richter12}
\bibinfo{author}{\bibfnamefont{A.}~\bibnamefont{Richter}},
  \bibinfo{author}{\bibfnamefont{S.W.}~\bibnamefont{Glunz}},
  \bibinfo{author}{\bibfnamefont{F.}~\bibnamefont{Werner}}, 
  \bibinfo{author} {\bibnamefont{J.~Schmidt}},  \bibnamefont{and} 
  \bibinfo{author}{\bibfnamefont{A.}~\bibnamefont{Cuevas}},
  \bibinfo{journal}{Phys. Rev. B} \textbf{\bibinfo{volume}{86}},
  \bibinfo{pages}{165202} (\bibinfo{year}{2012}).

\bibitem[{\citenamefont{J.Dziewior and Schmid}(1977)}]{dzie1977}
\bibinfo{author}{\bibnamefont{J.Dziewior}} \bibnamefont{and}
  \bibinfo{author}{\bibfnamefont{W.}~\bibnamefont{Schmid}},
  \bibinfo{journal}{Appl. Phys. Lett.} \textbf{\bibinfo{volume}{31}},
  \bibinfo{pages}{346} (\bibinfo{year}{1977}).

\bibitem[{\citenamefont{Lee and Gonze}(1995)}]{lee95}
\bibinfo{author}{\bibfnamefont{C.}~\bibnamefont{Lee}} \bibnamefont{and}
  \bibinfo{author}{\bibfnamefont{X.}~\bibnamefont{Gonze}},
  \bibinfo{journal}{Phys. Rev. B} \textbf{\bibinfo{volume}{51}},
  \bibinfo{pages}{8610} (\bibinfo{year}{1995}).

\bibitem[{\citenamefont{Takagi}(1958)}]{takagi58}
\bibinfo{author}{\bibfnamefont{S.}~\bibnamefont{Takagi}}, \bibinfo{journal}{J.
  Phys. Soc. Japan} \textbf{\bibinfo{volume}{13}}, \bibinfo{pages}{278}
  (\bibinfo{year}{1958}).

\bibitem[{\citenamefont{Thomas and E.Levine}(1965)}]{thomas1965}
\bibinfo{author}{\bibfnamefont{G.}~\bibnamefont{Thomas}} \bibnamefont{and}
  \bibinfo{author}{\bibnamefont{E.Levine}}, \bibinfo{journal}{Phys. Stat. Sol}
  \textbf{\bibinfo{volume}{11}}, \bibinfo{pages}{81} (\bibinfo{year}{1965}).

\bibitem[{\citenamefont{Howie and Whelan}(1961)}]{howie61}
\bibinfo{author}{\bibfnamefont{A.}~\bibnamefont{Howie}} \bibnamefont{and}
  \bibinfo{author}{\bibfnamefont{M.~J.} \bibnamefont{Whelan}},
  \bibinfo{journal}{Proc. R. Soc. Lond. A} \textbf{\bibinfo{volume}{263}},
  \bibinfo{pages}{217} (\bibinfo{year}{1961}).

\bibitem[{\citenamefont{Stadelmann}(2004)}]{JEMS}
\bibinfo{author}{\bibfnamefont{P.~A.} \bibnamefont{Stadelmann}},
  \emph{\bibinfo{title}{JEMS-EMS java version}} (\bibinfo{year}{2004}),
  \urlprefix\url{www.jems-saas.ch}.

\end{thebibliography}

\end{document}

\title{Observation of large multiple scattering effects\\ 
in ultrafast electron diffraction on single crystal silicon}

\author{I. Gonzalez Vallejo,$^{1,2}$ G. Gall\'e,$^1$ B. Arnaud,$^3$ S.~A. Scott,$^4$ M.~G. Lagally,$^4$ D. Boschetto,$^1$ P.-E. Coulon,$^5$ G. Rizza,$^5$ F. Houdellier,$^6$ D. Le Bolloc'h,$^2$ and J. Faure$^1$}

\affiliation{$^1$LOA, ENSTA ParisTech, CNRS, Ecole polytechnique, Univ. Paris-Saclay, Palaiseau, France}
\affiliation{$^2$LPS, CNRS, Univ. Paris-Sud, Univ. Paris-Saclay, 91405 Orsay, France}
\affiliation{$^3$Institut des Mol\'ecules et Mat\'eriaux du Mans, UMR CNRS 6283, Le Mans univ. , 72085 Le Mans, France}
\affiliation{$^4$University of Wisconsin-Madison, Madison, Wisconsin 53706, USA}
\affiliation{$^5$LSI, CNRS, CEA-DRF-IRAMIS, Ecole Polytechnique, Univ. Paris-Saclay, Palaiseau, France}
\affiliation{$^6$CEMES, CNRS, 29 Rue Jeanne Marvig, 31055 Toulouse, France}

\maketitle

 \begin{figure}[t!]
\centerline{\includegraphics[width=14cm]{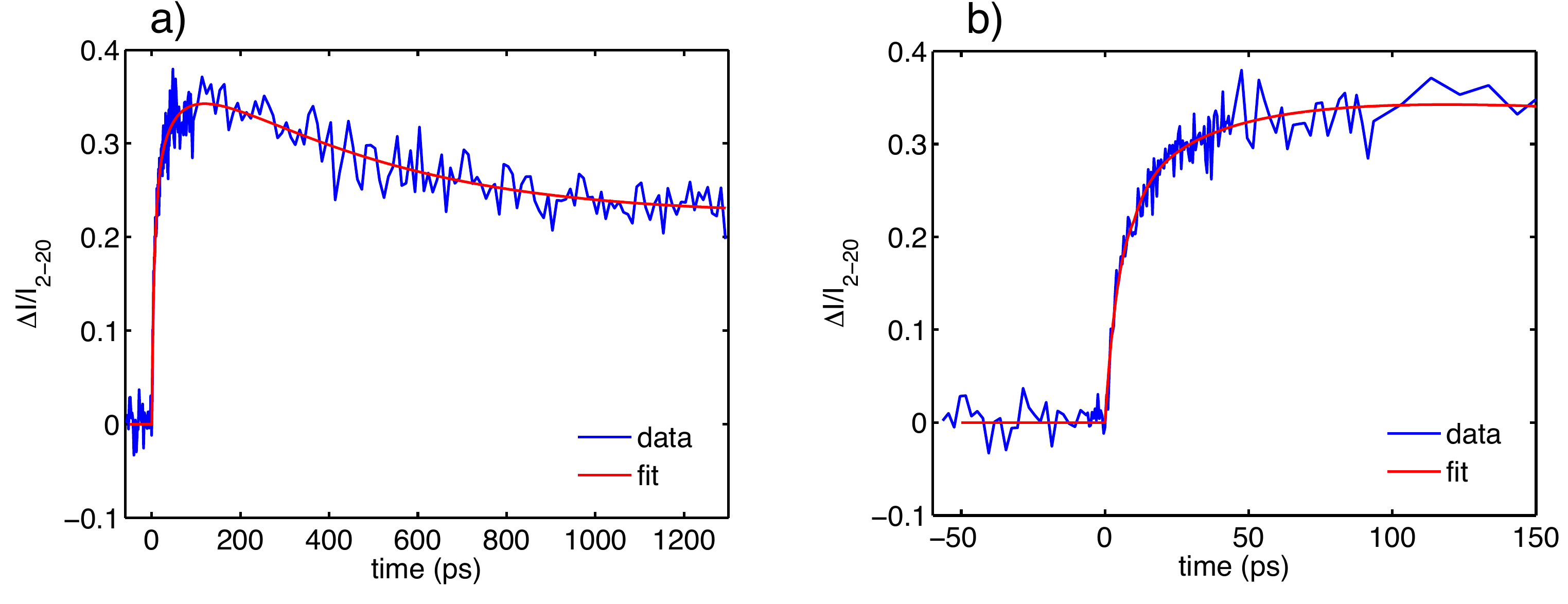}}
\caption{Relative intensity changes of the (2-20) Bragg peak as a function of time. Blue curve: data, red curve: fit. The pump fluence for this data was $8.2\mJcm$ and the electron energy was $45\kev$. Left: dynamics on the nanosecond time scale. Right: zoom on the $100\ps$ time scale.}\label{figS1}
\end{figure}

\section{Bragg peak dynamics on nanosecond time scale}
In this section, we give more information on the long time dynamics of the Bragg peaks. Figure~\ref{figS1} shows the relative intensity change $\Delta I(t)/I$ of the (2-20) Bragg peak for a pump fluence of $8.2\mJcm$ and a probing electron energy of $45\kev$. The sample was oriented so that the Bragg condition is fulfilled for the (2-20) Bragg peak. In order to extract the different time scales, we use the following phenomenological function for fitting the data
$$
f(t)=A_1(1-e^{-t/\tau_1})H(t) + A_2\left( 1-\frac{1}{\sqrt{1+(t-t_0)/\tau_2}} \right)H(t-t_0)+A_3\left( e^{-t/\tau_3}-1 \right)H(t-t_0)
$$
where $H(t)$ is the heaviside distribution and the $A_i$ and the $\tau_i$ are the free parameters of the fit. The first term is used to model the fast time scale that is present at the beginning of the data, probably due to electron-phonon scattering. The second term takes the form of the solution of the Auger recombination equation: $dn_e/dt=-Cn_e^3$. It has typical timescale $\tau_2$ and it is responsible for delayed heating, as expected from Auger recombination. Finally the third term represents the response of the system on a longer time scale, such as cooling through thermal diffusion. Note that we impose that the second and third processes occur after the first process through the use of the heaviside $H(t-t_0)$ with $t_0=2\tau_1$. The fits follow quite closely the experimental curve, as shown by the red curves in fig.~\ref{figS1}. We find that the first process, electron-phonon coupling, has a typical time scale of $\tau_1=5\ps$, while the second process, Auger, occurs on $\tau_2=32\ps$ and the slower cooling process on $\tau_3=450\ps$.

\section{N-beam dynamical diffraction theory}
In dynamical theory,  the main electron beam diffracts into $N-1$ diffracted beams because of its interaction with the crystal scattering potentiel. The crystal scattering potential is developed into a Fourier series as
\begin{equation}
V(\mathbf{r})=\sum_\mathbf{g}U_\mathbf{g}e^{i\mathbf{g}\cdot\mathbf{r}}
\end{equation}
where $\mathbf{g}$ are the lattice reciprocal vectors, and $U_\mathbf{g}$ are the potential Fourier components corresponding to $\mathbf{g}$. The scattered wave function is also written as a Fourier series:  $|\psi\rangle=\sum_\mathbf{g}\phi_\mathbf{g}|\mathbf{k}+\mathbf{g}\rangle$, where $\phi_\mathbf{g}$ are the amplitude of the scattered wave in diffraction peak corresponding to vector $\mathbf{g}$. Injecting these expressions into the Shrödinger equation and solving in Fourier space, one obtains the Howie-Whelan equations:
\begin{equation}
\frac{\partial\phi_{\mathbf{g}}}{\partial z}=is_{\mathbf{g}}\phi_{\mathbf{g}}+\sum_{\mathbf{g'}\neq\mathbf{g}}\frac{i}{2\xi_{\mathbf{g-g'}}}\phi_{\mathbf{g'}}
\end{equation}
where $s_\mathbf{g}$, the deviation error, depends on the crystal orientation, and $\xi_{\mathbf{g}-\mathbf{g'}}=\frac{1}{\lambda}\frac{2\hbar^2}{m_eU_g}$ is the extinction distance. The extinction distance is related to $U_{\mathbf{g}-\mathbf{g'}}$ which causes a coupling of the two diffracted beams $\phi_{\mathbf{g}}$ and $\phi_{\mathbf{g'}}$ because $\langle\mathbf{k}+\mathbf{g}|\hat V| \mathbf{k}+\mathbf{g'}\rangle=\langle\mathbf{k}+\mathbf{g}|U_{\mathbf{g}-\mathbf{g'}}e^{i(\mathbf{g}-\mathbf{g'})}| \mathbf{k}+\mathbf{g'}\rangle\neq0$. Note that for a weakly relativistic electron, the effect of the relativistic mass increase can be included simply by replacing $m_e$ by $\gamma m_e$ where $\gamma=1+E/m_ec^2$ is the electron Lorentz factor. The Howie-Whelan equations describe the evolution of the scattered wave intensities during propagation of the electron into the sample. This system of N coupled differential equations can be written in matrix form:
\begin{equation}
\frac{d\Phi}{dz}=iM\Phi
\end{equation}
where $\Phi$ is a column vector of length $N$ and $M$ is a $N\times N$ matrix that can be decomposed as
$$
M= \left( {\begin{array}{cccc}
   0 & 0 & 0 & \cdots \\
   0 & s_{g_1} & 0 &  \cdots \\
0 & 0 & s_{g_2} & \cdots\\
\vdots & \vdots & & \ddots  \\
  \end{array} } \right)+\frac{ \gamma m_e}{2\pi\hbar^2}\lambda
  \left( {\begin{array}{ccccc}
  0 & U_{-g_1} & U_{-g_2} & U_{-g_3} & \cdots \\
   U_{g_1} & 0 & U_{g_1-g_2}& U_{g_1-g_3} & \cdots \\
U_{g_2} & U_{g_2-g_1} & 0 & U_{g_2-g_3} & \cdots\\
\vdots & \vdots &  & \ddots  &  \\
  \end{array} } \right)
$$
The left matrix is diagonal and its elements are the amplitudes of the deviation vectors for each diffraction peak $s_{\mathbf{g}-\mathbf{g}'}$. The right matrix is composed of the Fourier amplitudes $U_{\mathbf{g}-\mathbf{g}'}.$ This is an eigenvalue problem and the solution is found by diagonalizing matrix $M$. If $D$ is the diagonal matrix in the basis of eigenvectors and $C$ is the matrix for changing basis, we have $M=CDC^{-1}$ and the solution of the problem is given by
$$
\Phi(z)=Ce^{iDz}C^{-1}\Phi(0)
$$

\begin{figure}[t]
\centerline{\includegraphics[width=8cm]{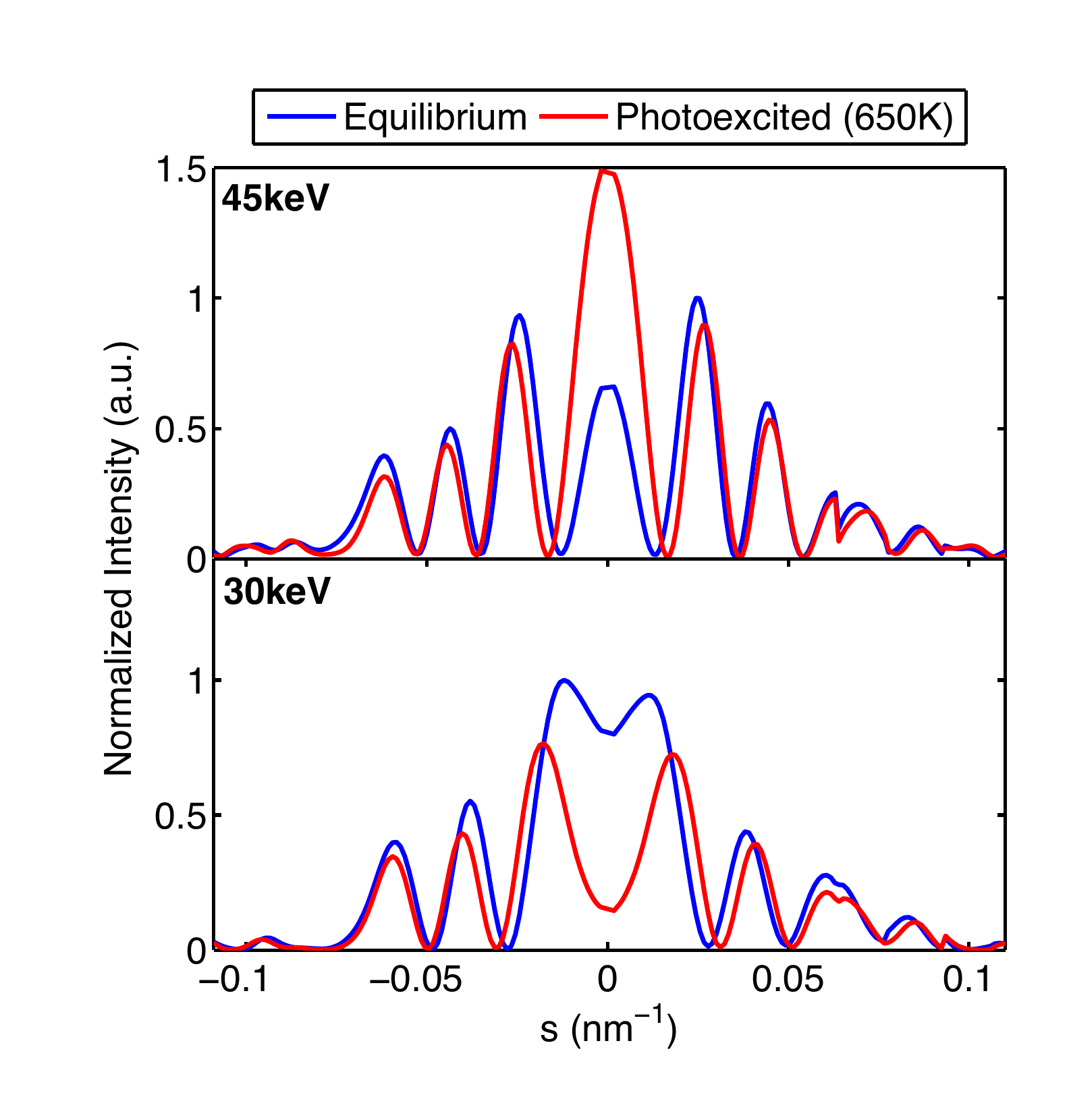}}
\caption{Results of dynamical diffraction theory including $N=26$ beams. We assume a $70\nm$ silicon sample probed by $45\kev$ electrons (top) and $30\kev$ electrons (bottom) for two different temperature, $300\K$ (blue curves) and $650\K$ (red curves).}\label{figS2}
\end{figure}

This general solution allows us to compute the amplitude of the various diffracted peaks $\phi_{\mathbf{g}}(L)$ at the output of the crystal, $z=L$. This theory can be used provided that the crystal potential $V(r)$ is precisely known. In our case,  we extracted the $U_\mathbf{g}$ matrix for the silicon potential from the code JEMS. We modeled the experiment assuming a $70\nm$ thickness and considering $N=26$ beams, including all (220), (400), (440), (620) peaks and a few higher order peaks as well. Such a high number of beams was necessary to ensure the convergence in the shape of the (220) rocking curve. Note that there are no free parameters in this model. Figure~\ref{figS2} shows the results of the calculations for $E=45\kev$ electrons and $30\kev$ electrons. The experimental trends are well reproduced: the shapes of the calculated rocking curves are similar to the experimental ones. In particular, the signs of the relative intensity change is reproduced: $\Delta I(s=0)/I>0$ at $45\kev$ and $\Delta I(s=0)/I<0$ at $30\kev$. 

However, the experimental data in fig.~3 in the main manuscript also displays a large background and the diffracted intensity $I(s)$ oscillates but never cancels to zero, in contradiction with dynamical diffraction calculations. Experimentally, the background can be due to many factors, such as inelastic scattering (on phonons, plasmons, defects...), surface contamination or surface amorphization. Because of the difficulty of modeling all these effects, we turn to a phenomenological approach and model the background using a simple gaussian distribution. We were able to obtain a quantitative fit of the experimental data using the following function:
$$
I(s)=AI_{dyn}+Be^{-s^2/\sigma^2}
$$
where $I_{dyn}$ is given by dynamical theory (no free parameters) and $A$, $B$ and $\sigma$ are free parameters allowing us to fit the experimental data more accurately.

\begin{figure}[t!]
\centerline{\includegraphics[width=12cm]{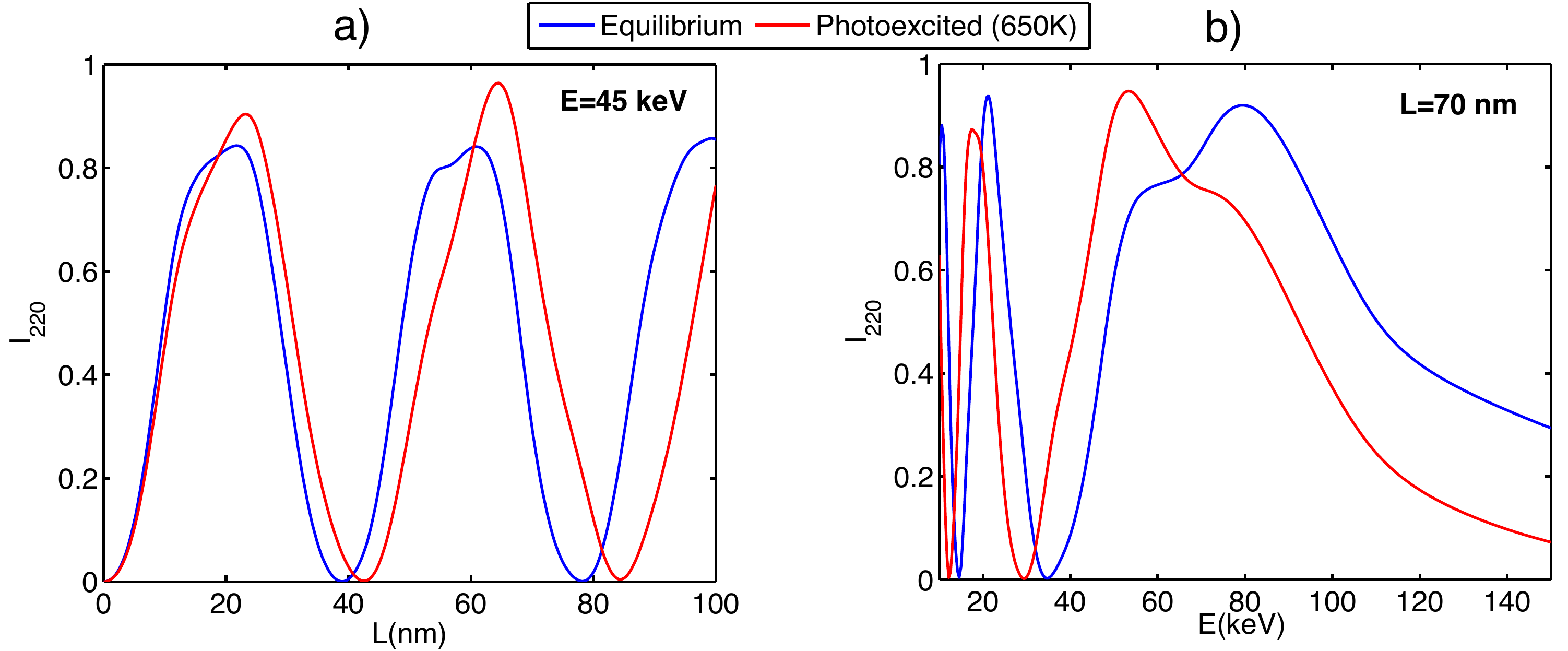}}
\caption{Results of dynamical diffraction theory including $N=26$ beams. a) Intensity of the (220) peak at the Bragg angle, $I_{220}(s=0)$ for varying sample thicknesses, assuming $45\kev$ electrons. b) Intensity of the (220) peak at the Bragg angle, $I_{220}(s=0)$ for varying electron energy, assuming a $70\nm$ thickness.}\label{figS3}
\end{figure}
As a complement, we show in fig.~\ref{figS3} different non intuitive behaviors of dynamical diffraction effects. Figure~\ref{figS3}a) shows the evolution of the (220) peak at the Bragg angle, $s=0$, as a function of thickness. The diffracted intensity oscillates along propagation in the sample which is one of the main feature of dynamical diffraction. Interestingly, the diffracted intensity in the high temperature case (red curve) shows a different behavior, indicating that the relative intensity changes are also expected to change sign depending on the sample thickness. Note that for small thicknesses, one recovers kinematical theory and $\Delta I/I<0$, i.e. the diffracted intensity is smaller in the high temperature case. Figure~\ref{figS3}b) shows a similarly complex behavior when the electron energy is varied. This indicates that the relative intensity $\Delta I/I$ have varying amplitude and sign depending on the energy of the probing electrons. We conclude that the ultrafast response of the Bragg peak intensity is, in general, greatly dependent on the sample thickness and the electron energy.

%\bibliographystyle{apsrev}
%\bibliography{jerome_0813,jerome_SS_0813,Bib_Benoit}
%
%
%%\begin{thebibliography}{39}
%\expandafter\ifx\csname natexlab\endcsname\relax\def\natexlab#1{#1}\fi
%\expandafter\ifx\csname bibnamefont\endcsname\relax
%  \def\bibnamefont#1{#1}\fi
%\expandafter\ifx\csname bibfnamefont\endcsname\relax
%  \def\bibfnamefont#1{#1}\fi
%\expandafter\ifx\csname citenamefont\endcsname\relax
%  \def\citenamefont#1{#1}\fi
%\expandafter\ifx\csname url\endcsname\relax
%  \def\url#1{\texttt{#1}}\fi
%\expandafter\ifx\csname urlprefix\endcsname\relax\def\urlprefix{URL }\fi
%\providecommand{\bibinfo}[2]{#2}
%\providecommand{\eprint}[2][]{\url{#2}}
%
%
%\bibitem[{\citenamefont{Brantov et~al.}(2008)\citenamefont{Brantov, Esirkepov,
%  Kando, Kotaki, Bychenkov, and Bulanov}}]{bran08}
%\bibinfo{author}{\bibfnamefont{A.~V.} \bibnamefont{Brantov}},
%  \bibinfo{author}{\bibfnamefont{T.~Z.} \bibnamefont{Esirkepov}},
%  \bibinfo{author}{\bibfnamefont{M.}~\bibnamefont{Kando}},
%  \bibinfo{author}{\bibfnamefont{H.}~\bibnamefont{Kotaki}},
%  \bibinfo{author}{\bibfnamefont{V.~Y.} \bibnamefont{Bychenkov}},
%  \bibnamefont{and} \bibinfo{author}{\bibfnamefont{S.~V.}
%  \bibnamefont{Bulanov}}, \bibinfo{journal}{Phys. Plasmas}
%  \textbf{\bibinfo{volume}{15}}, \bibinfo{eid}{073111} (\bibinfo{year}{2008}).
%
%
%\bibitem[{\citenamefont{Li13}(2013)}]{li13}
%\bibinfo{author}{\bibnamefont{F. Y. Li}} \bibnamefont{et}
%  \bibinfo{author}{\bibnamefont{al}},
%  \bibinfo{journal}{Phys. Rev. Lett.} \textbf{\bibinfo{volume}{110}},
%  \bibinfo{pages}{135002} (\bibinfo{year}{2013}).
%
%\end{thebibliography}
%